\documentclass[11.5pt, letterpaper]{amsart}
\usepackage[left=0.8in,right=0.8in,bottom=0.8in,top=0.8in]{geometry}
\usepackage{amsfonts}
\usepackage{amsmath,amssymb}
\usepackage{graphicx}
\usepackage[font=small,labelfont=bf]{caption}
\usepackage{epstopdf}
\usepackage[pdfpagelabels,hyperindex]{hyperref }
\usepackage{xcolor} 
\usepackage{amsthm}
\usepackage{float}
\usepackage{pgfplots}
\usepackage{listings} 
\usepackage{longtable}
\usepackage{mathrsfs}

\hypersetup{
pdftitle={The Anisotropic Gaussian Semi-Classical Schr\"{o}dinger Propagator},
pdfauthor={Panos D Karageorge},
pdfauthor={George N Makrakis},
pdfkeywords={Schr\"{o}dinger Equation, Schr\"{o}dinger Propagator, Semi-Classical Gaussian Wave Packets, Variational System, Riccati Equation, Van Vleck Formula.}
}

\newtheorem{thm}{Theorem}[section]

\newtheorem{prop}[thm]{Proposition}

\theoremstyle{definition}
\newtheorem{defn}[thm]{Definition}

\theoremstyle{remark}
\newtheorem{rem}[thm]{Remark}

\theoremstyle{remark}
\newtheorem{rmk}{Remark}

\makeatletter
\let\c@equation\c@thm
\makeatother
\numberwithin{equation}{section}

\author[Panos D Karageorge]{Panos D Karageorge$^1$}
\address{$^1$Department of Mathematics and Applied Mathematics, University of Crete, Voutes Campus, 700 13 Heraklion, Greece}
\email{pkarag@uoc.gr}

\author[George N Makrakis]{George N Makrakis$^2$}
\address{$^2$Department of Mathematics and Applied Mathematics, University of Crete, Voutes Campus, 700 13 Heraklion, Greece and Institute of Applied and Computational Mathematics, Foundation for Research and Technology, 100 Nikolaou Plastira St, Vassilika Vouton, 700 13 Heraklion, Greece}
\email{\texttt{makrakg}@\texttt{uoc.gr} \texttt{g.n.makrakis}@\texttt{iacm.forth.gr}}

\title{The Anisotropic Gaussian Semi-Classical Schr\"{o}dinger Propagator}
\begin{document}

\begin{abstract}
We present a construction of the Anisotropic Gaussian Semi-Classical Schr\"{o}dinger Propagator, emblematic of a class of Fourier Integral Operators of quadratic phase kernels related to the Schr\"{o}dinger equation. We deduce a set of algebraic relations of the variational matrices, solutions of the variational system pertaining to single Gaussian wave packet semi-classical time evolution, representing the symplectic and other invariances of the dynamics, which are subsequently used to derive the Van Vleck formula from the semi-classical propagator, as an argument for the practical importance of the later relations in the relevant wave packet calculus.
\end{abstract}

\maketitle

\centerline{\today}

The \textit{Anisotropic Gaussian Approximation} has been studied in the Physics literature in relation to \textit{Initial Value Representations} \cite{HeKl}, in the context of semi-classical time evolution problems in Atomic Physics and Theoretical Chemistry. The initial focus of the methodology was sinlge semi-classical Gaussian wave packet dynamics as an approximation of time evolution of more general quantum states \cite{Hel,HHH,HHL}. Semi-Classical Gaussian wave packet dynamics has been further advanced in the setting of general classes of problems of interest, for different conventions for the definition of semi-classical Gaussian wave packets and the operators generating the dynamics with respect to choice of quantization, and there have been detailed analyses of the their accuracy \cite{Lit,Hag,BBT,NSS,Rob1}. Similar ideas have been implemented in the construction of asymptotic solutions for the Cauchy problem of the wave equation, referred to as \textit{Gaussian beams} \cite{BaDa,Ral,KKL,CoFe}.

More recently, Fourier Integral Operators of quadratic phase kernels related to the Schr\"{o}dinger equation have been studied in the Mathematics literature, with complex valued quadratic phases of their kernels and common algebraic and geometric structure underlying their dynamics \cite{LaSi,PaUr,Rob2,RoSw}. Insofar as the underlying algebraic structure, a relation between the underlying Riccati dynamics and the Nearby Orbit method has been established, in essence a relation between the anisotropy matrix controlling the direction and spread of the propagated wave packet, a solution of the matrix Riccati equation, and the variational matrices, solutions of the corresponding variational system \cite{Zel,Hag}.

In this paper, we consider the problem of semi-classical time evolution of single semi-classical Gaussian wave packets for the Weyl quantization, for a general class of time dependent problems, as a basis to construct the \textit{Anisotropic Gaussian Semi-Classical Schr\"{o}dinger Propagator.} We derive a set of relations between the variational matrices, which appear as representations of the symplectic invariance of the underlying Hamiltonian flow and the symmetry of the Riccati dynamics, of particular practical use in the relevant calculus. We use these relations in order to derive the semi-classical Van Vleck formula from the constructed semi-classical propagator \cite{BiRo,Bla}.

\section{The Anisotropic Gaussian Approximation}

We begin with a constructive definition of the semi-classical Gaussian wave packet.

\bigskip

\begin{defn}
\textit{The} semi-classical Gaussian wave packet \textit{in $L^2(\mathbb{R}^d,dx;\mathbb{C})$, centered at the phase space point $(q,p)\in\mathbb{R}^{2d}$, is defined by the action of the Weyl shift \cite{Lit,Rob1,Fol}} 
\begin{equation}
\mathcal{T}_{(q,p)}=\exp\frac{i}{\hbar}\Big(p\cdot x-q\cdot \Big(-i\hbar\,\frac{\partial }{\partial x}\Big)\Big)
\end{equation}
\textit{on the Gaussian vacuum state, $G_0(x;\hbar)=(\pi\hbar)^{-d/4}e^{-|x|^2/2\hbar}$, i.e.,}
\begin{equation}
G_{(q,p)}(x;\hbar):=\mathcal{T}_{(q,p)}G_0(x;\hbar) \ .
\end{equation}
\end{defn}

\bigskip

\begin{prop}
\textit{According to the above definition, we have}
\begin{equation}
G_{(q,p)}(x;\hbar)=(\pi\hbar)^{-d/4}\,\exp\frac{i}{\hbar}\Big(\frac{p\cdot q}{2}+p\cdot(x-q)+\frac{i}{2}|x-q|^2\Big) \ .
\end{equation}
\textit{The set of semi-classical Gaussian wave packets, $\{G_{(q,p)}\}_{(q,p)\in\mathbb{R}^{2d}}$, constitutes a tight continuous frame in $L^2(\mathbb{R}^d)$.}
\end{prop}

\begin{proof}
The action of the Weyl shift is understood by means of the Baker-Campbell-Hausdorff formula for the Heisenberg algebra $\mathfrak{h}_{2d+1}$. Elements $h,h'\in \mathfrak{h}_{2d+1}$ are uniquely expressed as $h=a\cdot X+b\cdot Y+c\,E$ and $h'=a'\cdot X+b'\cdot Y+c'\,E$, for $(a,b,c),(a',b',c')\in\mathbb{R}^{2d+1}$, where $X=(X_1,\ldots,X_d)$, $Y=(Y_1,\ldots,Y_d)$ and $E$ are generators of the algebra satisfying the commutation relations $[X_j,Y_k]=\delta_{jk}\,E$, $[X_j,E]=0$ and $[Y_j,E]=0$, for $j,k=1,\ldots,d$ \cite{Lit,Fol}. For the exponential map onto the Heisenberg group $\mathbb{H}_{2d+1}$, we have
\begin{equation}
e^{h}e^{h'}=\exp\Big(h+h'+\frac{1}{2}[h,h']\Big)
\end{equation}
so that unambiguous meaning is given to the two non-commuting argument function 
\begin{equation}
e^{h+h'}=\exp\Big(-\frac{1}{2}(a'\cdot b-a\cdot b')\,E\Big)e^{h}e^{h'} \ .
\end{equation}
The action of the Weyl shift follows, as it realizes the Schr\"{o}dinger representation of $\mathbb{H}_{2d+1}$ on $L^2(\mathbb{R}^d)$,
\begin{eqnarray}
\mathcal{T}_{(q,p)}G_0(x;\hbar)=\exp\Bigg(-\frac{1}{2}\Big[\frac{i}{\hbar}\,p\cdot x,-\frac{i}{\hbar}\,q\cdot \Big(-i\hbar\, \frac{\partial }{\partial x}\Big)\Big]\Bigg)e^{\frac{i}{\hbar}\,p\cdot x}\exp\Bigg(-\frac{i}{\hbar}\,q\cdot \Big(-i\hbar\,\frac{\partial }{\partial x}\Big)\Bigg)G_0(x;\hbar)\\ \nonumber
=\exp\frac{i}{\hbar}\Big(p\cdot x-\frac{p\cdot q}{2}\Big)\,\exp\Big(-q\cdot \frac{\partial }{\partial x}\Big)G_0(x;\hbar)=(\pi\hbar)^{-d/4}\,\exp\frac{i}{\hbar}\Big(\frac{p\cdot q}{2}+p\cdot(x-q)+\frac{i}{2}|x-q|^2\Big)
\end{eqnarray}
noting that $[p\cdot x,q\cdot \frac{\partial}{\partial x}]=-p\cdot q\,{\rm Id}$.

As for the second point, the set of semi-classical Gaussian wave packets defines a map from the phase space, as a measure space endowed with the Liouville measure, to $L^2(\mathbb{R}^d)$, $(q,p)\mapsto G_{(q,p)}$, such that \cite{TAG,RDD}: 

1. it is weakly measurable, in the sense that for all $\psi\in L^2(\mathbb{R}^d)$, the function $\langle G_{(q,p)},\psi\rangle $ is measurable with respect to the Liouville measure; the later is proportional to the \textit{phase space wave function,} which possesses strong smoothness properties\cite{KaMa},

2. for all $\psi\in L^2(\mathbb{R}^d)$ the inequalities $c_1\,\|\psi\|_{L^2}^2\leq \int|\langle G_{(q,p)},\psi\rangle|^2 \,dqdp\leq c_1\,\|\psi\|_{L^2}^2$ are saturated for $c_1=c_2=(2\pi\hbar)^{d}$ \cite{KaMa}.

\end{proof}

\bigskip

\begin{rem}
Some authors use alternative definitions for semi-classical Gaussian wave packets, all of which share the form 
\begin{equation}
\psi_{(q,p)}(x;\hbar)=(\pi\hbar)^{-d/4}\,\exp\frac{i}{\hbar}\Big(\phi(q,p)+p\cdot (x-q)+\frac{i}{2}|x-q|^2\Big)
\end{equation} 
where the real phase $\phi$ satisfies the condition $\phi(0)=0$ \cite{Fol}.
\end{rem}

\bigskip

The continuous frame $\{G_{(q,p)}\}_{(q,p)\in\mathbb{R}^{2d}}$ furnishes a phase space resolution for the Schr\"{o}dinger propagator \cite{TAG,Rob1}
\begin{equation}
U(t_0,t)=\Big(\frac{1}{2\pi\hbar}\Big)^{d}\int U(t_0,t) G_{(q,p)}\langle G_{(q,p)},\cdot\rangle \,dqdp
\end{equation}
generated by the Weyl quantization $\widehat H$ of a Hamiltonian function of appropriate smoothness and growth, which is the basis for the \textit{Initial Value Representations} for semi-classical Schr\"{o}dinger propagators \cite{HeKl,Lit}. 

Setting off from the above resolution, one constructs a semi-classical Schr\"{o}dinger propagator by the \textit{Anisotropic Gaussian Approximation.} This amounts to super-posing asymptotic solutions of the semi-classical Cauchy problem for the Schr\"{o}dinger equation
\begin{equation}
i\hbar\,\frac{\partial\psi}{\partial t}=\widehat H\psi \ , \ \ t\in[t_0,t_0+T] \ , \ \ \psi(t_0)=G_{( q, p)}
\label{eq:cauchyproblem}
\end{equation}
by an ansatz closely related to the \textit{Nearby Orbit Approximation,} i.e., linearization of the Hamiltonian flow at an orbit emanating from $(q,p)$ \cite{Hel,HHH,Lit,NSS,BBT,Rob1}.

\bigskip

\begin{defn} 
\textit{The} Variational System \textit{related to the Nearby Orbit Approximation for the Hamiltonian flow generated by a Hamiltonian function $H$, smooth in $(q,p)\in\mathbb{R}^{2d}$ and continuous in $t\in\mathbb{R}$, is as follows}
\begin{equation}
\frac{d}{dt}\left(
\begin{array}{ccc}
A \\
B
\end{array}\right)=\left(\begin{array}{ccc}
H_{pq} & H_{pp} \\
-H_{qq} & -H_{qp} 
\end{array}\right)\left(
\begin{array}{ccc}
A \\
B 
\end{array}\right) 
\end{equation}
\textit{the constituent Hessian blocks evaluated along the orbit $(q_t,p_t)=g^t(q,p)$}
\begin{equation}
\frac{d}{dt}\left(
\begin{array}{ccc}
A \\
B
\end{array}\right)=J\,{\rm Hess}_{(q,p)}\,H(q_t,p_t,t)\left(
\begin{array}{ccc}
A \\
B 
\end{array}\right)
\end{equation}
\textit{where $J=\left(\begin{array}{ccc} 0 & I \\ -I & 0 \end{array} \right)$ is the $2d\times 2d$ canonical symplectic matrix, the accompanying initial conditions being $A(t_0)=I$ and $B(t_0)=iI$, where $I$ is the $d\times d$ identity matrix. The solutions, $A$ and $B$, are termed the} position \textit{and} momentum variational matrix, \textit{respectively.}
\end{defn}

\bigskip

\begin{prop}\textit{The solution of the initial value problem for the variational system has the dynamical representation \cite{NSS,Rob1}} 
\begin{equation}
A(q,p,t_0,t)=\frac{\partial q_t}{\partial q}+i\frac{\partial q_t}{\partial p} \ , \ \ B(q,p,t_0,t)=\frac{\partial p_t}{\partial q}+i\frac{\partial p_t}{\partial p} \ .
\label{eq:vmatrices}
\end{equation}
The fundamental solution of the variational system, 
defined as the solution of the initial value problem
\begin{equation}
\frac{d\sigma}{dt}=J\,{\rm Hess}_{(q,p)}\,H(q_t,p_t,t)\,\sigma \ , \ \ \sigma(t_0)=I
\end{equation}
where $I$ is the $2d\times 2d$ identity matrix, admits the formal time ordered exponential series 
\begin{equation}
\sigma(q,p,t_0,t)=\mathcal{T}{\rm exp}\int\displaylimits_{t_0}^tJ\,{\rm Hess}_{(q,p)}\,H(q_\tau,p_\tau,\tau)\,d\tau
\end{equation}
and is given by the (symplectic) Jacobian matrix of the Hamiltonian flow 
\begin{equation}
\sigma(q,p,t_0,t)=\frac{\partial(q_t,p_t)}{\partial (q,p)} \ .
\end{equation}
The variational matrices are expressed by means of the fundamental solution as follows 
\begin{equation}
\left(
\begin{array}{ccc}
A(t) \\
B(t)
\end{array}\right)=\sigma(t)\left(
\begin{array}{ccc}
I \\
iI 
\end{array}\right) \ .
\end{equation}

\end{prop}

\begin{proof}
By virtue of Hamilton's equations 
\begin{equation}
\frac{dq_t}{dt}=\frac{\partial H}{\partial p}(q_t,p_t,t) \ , \ \ \frac{dp_t}{dt}=-\frac{\partial H}{\partial q}(q_t,p_t,t)
\end{equation} 
with initial conditions $q_{t_0}=q$ and $p_{t_0}=p$, and as $H$ is smooth in $(q,p)\in\mathbb{R}^{2d}$, we have 
\begin{eqnarray}
\frac{d}{dt}\Big(\frac{\partial q_t}{\partial q}+i\frac{\partial q_t}{\partial p}\Big)=\frac{\partial}{\partial q}\frac{dq_t}{dt}+i\frac{\partial}{\partial p}\frac{dq_t}{dt}=\frac{\partial}{\partial q}\Big(\frac{\partial H}{\partial p}(q_t,p_t,t)\Big)+i\frac{\partial}{\partial p}\Big(\frac{\partial H}{\partial p}(q_t,p_t,t)\Big)\\ \nonumber 
=H_{pq}(q_t,p_t,t)\Big(\frac{\partial q_t}{\partial q}+i\frac{\partial q_t}{\partial p}\Big)+H_{pp}(q_t,p_t,t)\Big(\frac{\partial p_t}{\partial q}+i\frac{\partial p_t}{\partial p}\Big)
\end{eqnarray}
while we similarly approach $\frac{d}{dt}\Big(\frac{\partial p_t}{\partial q}+i\frac{\partial p_t}{\partial p}\Big)$; noting, additionally, that 
\begin{equation}
\Big(\frac{\partial q_t}{\partial q}+i\frac{\partial q_t}{\partial p}\Big)\Big|_{t=t_0}=I \ , \ \ \Big(\frac{\partial p_t}{\partial q}+i\frac{\partial p_t}{\partial p}\Big)\Big|_{t=t_0}=iI
\end{equation}
by uniqueness of the solution of the initial value problem for the variational system, we arrive at the conclusion. The construction of the fundamental solutions follows by similar arguments.
\end{proof}

\bigskip

We proceed to the issue of asymptotic solution of the semi-classical Cauchy problem for the Schr\"{o}dinger equation.

\bigskip

\begin{thm}\label{charsys} \textit{Let $t_0\in\mathbb{R}$ be a time instant, $T>0$ a fixed time interval, and $H$ a Hamiltonian function, smooth in $(q,p)\in\mathbb{R}^{2d}$ and continuous in $t\in\mathbb{R}$, such that for any $\alpha,\beta\in\mathbb{N}_0^d$ there exists some $c_{\alpha\beta}(t_0,T)>0$ and $\ell_{|\alpha+\beta|}(t_0,T)\in\mathbb{R}$, such that, for any $(q,p)\in\mathbb{R}^{2d}$ and $t\in[t_0,t_0+T]$, the following estimate holds \cite{Rob1,NSS}}
\begin{equation}
\Big|\frac{\partial^{\alpha+\beta}H}{\partial q^\alpha\partial p^\beta}(q,p,t)\Big|\leq c_{\alpha\beta}(t_0,T)\Big(1+|q|+|p|\Big)^{\ell_{|\alpha+\beta|}(t_0,T)} \ .
\label{eq:estimate}
\end{equation}
\textit{The} Anisotropic Gaussian Approximation \textit{for the semi-classical Cauchy problem (\ref{eq:cauchyproblem}), for given $(q,p)\in\mathbb{R}^{2d}$, amounts to the anisotropic semi-classical Gaussian wave packet being an asymptotic solution}
\begin{equation}
G_{( q, p)}^Z(x,t;\hbar)=(\pi\hbar)^{-d/4}a(t)\,\exp\frac{i}{\hbar}\Big(\frac{p\cdot q}{2}+S(t)+p_t\cdot(x-q_t)+\frac{1}{2}(x-q_t)\cdot Z(t)(x-q_t)\Big) \ .
\label{eq:agwp}
\end{equation}
\textit{In particular, if the parameters $(q_t,p_t)$, $S(t)$, $Z(t)$ and $a(t)$, for which we assume differentiability in time for $t\geq t_0$, satisfy the} characteristic system, \textit{namely}
\begin{eqnarray}\label{eq:charsyst}
\frac{dq_t}{dt}=\frac{\partial H}{\partial p} \ , \ \ \frac{dp_t}{dt}=-\frac{\partial H}{\partial q} \ , \ \ (q_{t_0},p_{t_0})=( q, p)\\ \nonumber
\frac{dS}{dt}=p_t\cdot \frac{dq_t}{dt}-H \ , \ \ S(t_0)=0 \\ \nonumber
\frac{dZ}{dt}+ZH_{pp}Z+ZH_{pq}+H_{qp}Z+H_{qq}=0 \ , \ \ Z(t_0)=iI \\ \nonumber
\frac{da}{dt}+\frac{1}{2}\,{\rm tr}\,\Big(H_{pp}Z+H_{pq}\Big)\,a=0 \ , \ \ a(t_0)=1
\end{eqnarray}
\textit{where the Hamiltonian and its derivatives are evaluated at $(q_t,p_t,t)$, whose solutions are:}

\textit{1. the image of the Hamiltonian flow generated by $H$}
\begin{equation} 
(q_t( q, p),p_t( q, p)):=g^t(q,p)
\end{equation}

\textit{2. the} action of the orbit emanating from $(q,p)\in\mathbb{R}^{2d}$
\begin{equation}
S(q,p,t_0,t)=\int\displaylimits_{t_0}^t\Big(p_\tau \cdot\frac{dq_\tau }{d\tau}-H(q_\tau ,p_\tau ,\tau)\Big)\,d\tau
\end{equation}

\textit{3. the} anisotropy matrix
\begin{equation}
Z(q,p,t_0,t)=\Big(\frac{\partial p_t}{\partial q}+i\frac{\partial p_t}{\partial p}\Big)\Big(\frac{\partial q_t}{\partial q}+i\frac{\partial q_t}{\partial p}\Big)^{-1}
\label{eq:Z}
\end{equation}

\textit{4. the} amplitude
\begin{equation}
a(q,p,t_0,t)=\frac{1}{\sqrt{{\rm det}\,\Big(\frac{\partial q_t}{\partial q}+i\frac{\partial q_t}{\partial p}\Big)}}
\label{eq:a}
\end{equation}
\textit{where the branch of the square root is that for which the initial value $a(t_0)=1$ is reached continuously, then (\ref{eq:agwp}) is an asymptotic solution of (\ref{eq:cauchyproblem}) in the sense that there exists some $C(q,p,t_0,T)>0$ such that}
\begin{equation}
\Big\|\Big(i\hbar\,\frac{\partial}{\partial t}-\widehat H\Big)G^Z_{(q,p)}(\cdot ,t;\hbar)\Big\|_{L^2}\leq C(q,p,t_0,T)\,\hbar^{3/2} \ .
\label{eq:asol}
\end{equation}
\end{thm}

\begin{proof} We begin by deriving the commutation formula for the action of $\widehat H$ on the anisotropic semi-classical Gaussian wave packet $G^Z_{(q,p)}(t)$, which is well defined by virtue of (\ref{eq:estimate}), and given by \cite{BeSh} 
\begin{equation}
\widehat HG^Z_{(q,p)}(x,t;\hbar)=\Big(\frac{1}{2\pi\hbar}\Big)^d\int e^{\frac{i}{\hbar}\xi\cdot (x-y)}H\Big(\frac{x+y}{2},\xi,t\Big)G^Z_{(q,p)}(y,t;\hbar)\,dyd\xi \ .
\end{equation}
By integrating over the microscopic phase space co-ordinates $(y,\xi)\mapsto (Q,P)=\Big(\frac{(x+y)/2-q_t}{\sqrt{\hbar}},\frac{\xi-p_t}{\sqrt{\hbar}}\Big)$, localizing the mean of the points $x$ and $y$ near the orbit's position in configuration space and the the momentum $\xi$ near the orbit's momentum in the dual space, we obtain
\begin{eqnarray}
\widehat HG^Z_{(q,p)}(x,t;\hbar)=\Bigg(\pi^{-d}\int H\Big(q_t+\sqrt{\hbar}\,Q,p_t+\sqrt{\hbar}\,P,t\Big)\,e^{\frac{2i}{\sqrt{\hbar}}P\cdot (x-q_t)}\\ \nonumber
\times \exp\Big(2iQ\cdot ZQ-2i(P+\frac{1}{\sqrt{\hbar}}Z (x-q_t))\cdot Q\Big)\,dQdP\Bigg)\,G^Z_{(q,p)}(x,t;\hbar) \ .
\end{eqnarray}
By Taylor expanding the Hamiltonian at the orbit $(q_t,p_t)$ to second order, we obtain the commutation formula 
\begin{eqnarray}
\widehat HG^Z_{(q,p)}(x,t;\hbar)=G^Z_{(q,p)}(x,t;\hbar)\Bigg(H+\frac{\partial H}{\partial q}\cdot (x-q_t)+\frac{\partial H}{\partial q}\cdot Z(x-q_t)\\ \nonumber
+\frac{1}{2}(x-q_t)\cdot \Big(ZH_{pp} Z+ZH_{pq} +H_{qp} Z+H_{qq} \Big)(x-q_t)-\frac{i\hbar}{2}\,{\rm tr}\,\Big(H_{pp} Z+H_{pq}\Big)\\ \nonumber 
+\pi^{-d}\int r_{(q_t,p_t)}(Q,P,t;\hbar)\exp\Big(2iQ\cdot Z(t) Q-2iP\cdot Q+2i\,\frac{x-q_t}{\sqrt{\hbar}}\cdot P-2i\,\frac{x-q_t}{\sqrt{\hbar}}\cdot Z(t)Q\Big)\,dQdP\Bigg)
\end{eqnarray}
where the Hamiltonian and its derivatives are evaluated at the arguments $(q_t,p_t,t)$ and $r_{(q_t,p_t)}(Q,P,t;\hbar)$ is the Taylor remainder
\begin{equation}
r_{(q_t,p_t)}(Q,P,t;\hbar)=\hbar^{3/2}\sum_{|\alpha|+|\beta|=3}\frac{1}{\alpha!\beta!}\frac{\partial^{\alpha+\beta}H}{\partial q^\alpha\partial p^\beta}(q_t+\lambda\sqrt{\hbar}\,Q,p_t+\lambda\sqrt{\hbar}\,P,t)\,Q^\alpha P^\beta
\end{equation}
for some $\lambda\in\,]0,1[\,$. 

We define the remainder of the Schr\"{o}dinger equation with respect to $G^Z_{(q,p)}(t)$ as
\begin{eqnarray}
R^Z_{(q,p)}(x,t;\hbar):=\Big(i\hbar\,\frac{\partial}{\partial t}-\widehat H\Big)G^Z_{(q,p)}(x,t;\hbar)\\ \nonumber
=\Bigg(-\frac{dS}{dt}+p_t\cdot \frac{dq_t}{dt}-H+\sqrt{\hbar}\,\Big(-\frac{dp_t}{dt}+Z\frac{d q_t}{dt}-\frac{\partial H}{\partial q}-Z\frac{\partial H}{\partial p}\Big)\cdot \frac{x-q_t}{\sqrt{\hbar}}\\ \nonumber
+\hbar\,\Bigg[i\,a^{-1}\frac{da}{dt}+\frac{i}{2}\,{\rm tr}\,\Big(H_{pp}Z+H_{qp}\Big)+\frac{1}{2}\frac{x-q_t}{\sqrt{\hbar}}\cdot \Big(\frac{dZ}{dt}+ZH_{pp}Z+ZH_{pq}+H_{qp}Z+H_{qq}\Big)\frac{x-q_t}{\sqrt{\hbar}}\Bigg]\Bigg)\,G^Z_{(q,p)}(x,t;\hbar)\\ \nonumber
-\Bigg(\pi^{-d}\int r_{(q_t,p_t)}(Q,P,t;\hbar)\exp\Big(2iQ\cdot Z(t) Q-2iP\cdot Q+2i\,\frac{x-q_t}{\sqrt{\hbar}}\cdot P-2i\,\frac{x-q_t}{\sqrt{\hbar}}\cdot Z(t)Q\Big)\,dQdP\Bigg)\,G^Z_{(q,p)}(x,t;\hbar) \ .
\label{eq:sys}
\end{eqnarray} 

Assuming the characteristic system $(\ref{eq:charsyst})$ is satisfied, and introducing the configuration space co-ordinate localized at the trajectory $x=q_t$, i.e., $x\mapsto \chi =\frac{x-q_t}{\sqrt{\hbar}}$, the remainder reduces to the form 
\begin{eqnarray}
R^Z_{(q,p)}(q_t+\sqrt{\hbar}\,\chi,t;\hbar)=-\pi^{-d}\hbar^{3/2}\sum_{|\alpha|+|\beta|=3}\frac{1}{\alpha!\beta!}\Bigg(\int\frac{\partial^{\alpha+\beta}H}{\partial q^\alpha\partial p^\beta}(q_t+\lambda\sqrt{\hbar}\,Q,p_t+\lambda\sqrt{\hbar}\,P,t)\,Q^\alpha P^\beta\\ \nonumber 
\times \exp\Big(2iQ\cdot Z(t) Q-2iP\cdot Q+2i\,\chi\cdot P-2i\,\chi\cdot Z(t)Q\Big)\,dQdP\Bigg)\,G^Z_{(q,p)}(q_t+\sqrt{\hbar}\,\chi,t;\hbar) \ .
\end{eqnarray}

We proceed to estimate the the norm of the remainder by a direct estimate of the resulting oscillatory integral
\begin{eqnarray}
\|R^Z_{(q,p)}(\cdot,t;\hbar)\|_{L^2}=\pi^{-5d/4}\hbar^{3/2}|a(t)|\Bigg(\sum_{|\alpha|+|\beta|=3,\,|\alpha'|+|\beta'|=3}\frac{1}{\alpha!\beta!\alpha'!\beta'!}\\ \nonumber
\times \int f_{\alpha\beta}(Q,P,t;\hbar)f_{\alpha'\beta'}(Q',P',t;\hbar)\,\exp\,2i\Big((Q\cdot ZQ-Q'\cdot \bar ZQ')-(P\cdot Q-P'\cdot Q')\Big)\\ \nonumber
\times \exp\Big(-\chi\cdot Z_2\chi+2i\Big((P-ZQ)-(P'-\bar ZQ')\Big)\cdot \chi\Big)\,d\chi dQdPdQ'dP'\Bigg)^{1/2}
\end{eqnarray}
where 
\begin{equation}
f_{\alpha\beta}(Q,P,t;\hbar)=\frac{\partial^{\alpha+\beta}H}{\partial q^\alpha\partial p^\beta}(q_t+\lambda\sqrt{\hbar}\,Q,p_t+\lambda\sqrt{\hbar}\,P,t)\,Q^\alpha P^\beta \ .
\end{equation}

By the Fubini theorem we may perform the integration with respect to $\chi$ 
\begin{eqnarray}
\int \exp\Big(-\chi\cdot Z_2\chi+2i\Big((P-ZQ)-(P'-\bar ZQ')\Big)\cdot \chi\Big)\,d\chi\\ \nonumber 
=\frac{\pi^{d/2}}{\sqrt{{\rm det}\,Z_2}}\,\exp\Big(-\Big((P-ZQ)-(P'-\bar ZQ')\Big)\cdot Z_2^{-1}\Big((P-ZQ)-(P'-\bar ZQ')\Big)\Big)
\end{eqnarray}
as $Z_2$ is invertible and positive definite, by virtue of relation (\ref{eq:sqrts}) which is proven in theorem (\ref{VMI}) subsequently, so that we obtain
\begin{eqnarray}
\|R^Z_{(q,p)}(\cdot,t;\hbar)\|_{L^2}=\pi^{-d}\hbar^{3/2}\Bigg(\sum_{|\alpha|+|\beta|=3,\,|\alpha'|+|\beta'|=3}\frac{1}{\alpha!\beta!\alpha'!\beta'!}\\ \nonumber
\times \int f_{\alpha\beta}(Q,P,t;\hbar)f_{\alpha'\beta'}(Q',P',t;\hbar)\,\exp\,2i\Big((Q\cdot ZQ-Q'\cdot \bar ZQ')-(P\cdot Q-P'\cdot Q')\Big)\\ \nonumber
\times \exp\Big(-\Big((P-ZQ)-(P'-\bar ZQ')\Big)\cdot Z_2^{-1}\Big((P-ZQ)-(P'-\bar ZQ')\Big)\Big)\,dQdPdQ'dP'\Bigg)^{1/2} \ .
\label{eq:norm1}
\end{eqnarray}
By the linear change of variables 
\begin{equation}
(Q,P,Q',P')\mapsto \kappa(Q,P,Q',P')=(u,v,u',v')=(Z_1Q-P,Z_2Q,Z_1Q'-P',-Z_2Q')
\end{equation}
where $|{\rm det}\,\kappa|=({\rm det}\,Z_2(t))^{-2}$, the norm becomes
\begin{eqnarray}
\|R^Z_{(q,p)}(\cdot,t;\hbar)\|_{L^2}=\hbar^{3/2}\,\frac{\pi^{-d}}{{\rm det}\,Z_2(t)}\,\Bigg(\sum_{|\alpha|+|\beta|=3,\,|\alpha'|+|\beta'|=3}\frac{1}{\alpha!\beta!\alpha'!\beta'!}\\ \nonumber
\times \int F_{\alpha\beta}(u,v,t;\hbar)F_{\alpha'\beta'}(u',v',t;\hbar)\,\exp\Big(-(u,v,u',v')\cdot M(t)(u,v,u',v')\Big)\,dudvdu'dv'\Bigg)^{1/2}
\label{eq:norm2}
\end{eqnarray}
where the amplitude is $F_{\alpha\beta}(t;\hbar)=f_{\alpha\beta}\circ \kappa^{-1}(t;\hbar)$ and the matrix of the quadratic form of the phase is 
\begin{equation}
M(t)=\left(\begin{array}{cccc} 
Z_2^{-1} & 0 & \frac{1}{2}Z_2^{-1} & \frac{i}{2}Z_2^{-1} \\
0 & Z_2^{-1} & \frac{i}{2}Z_2^{-1} & -\frac{1}{2}Z_2^{-1} \\
\frac{1}{2}Z_2^{-1} & \frac{i}{2}Z_2^{-1} & Z_2^{-1} & 0 \\
\frac{i}{2}Z_2^{-1} & -\frac{1}{2}Z_2^{-1} & 0 & Z_2^{-1} 
\end{array} \right) \ .
\end{equation}

The real part of the later 
\begin{equation}
{\rm Re}\,M(t)=\left(\begin{array}{cccc} 
Z_2^{-1} & 0 & \frac{1}{2}Z_2^{-1} & 0 \\
0 & Z_2^{-1} & 0 & -\frac{1}{2}Z_2^{-1} \\
\frac{1}{2}Z_2^{-1} & 0 & Z_2^{-1} & 0 \\
0 & -\frac{1}{2}Z_2^{-1} & 0 & Z_2^{-1} 
\end{array} \right)
\end{equation}
is positive definite, as the necessary and sufficient conditions for positive definiteness hold \cite{HoZh}, i.e., both its upper left $d\times d$ constituent block matrix $\left(\begin{array}{cccc} 
Z_2^{-1} & 0 \\
0 & Z_2^{-1} 
\end{array} \right)$ and the corresponding Schur complement 
\begin{eqnarray}
{\rm Re}\,M/ \left(\begin{array}{cccc} 
Z_2^{-1} & 0 \\
0 & Z_2^{-1} 
\end{array} \right):=\left(\begin{array}{cccc} 
Z_2^{-1} & 0 \\
0 & Z_2^{-1} 
\end{array} \right)\\ \nonumber
-\left(\begin{array}{cccc} 
\frac{1}{2}Z_2^{-1} & 0 \\
0 & -\frac{1}{2}Z_2^{-1}\end{array} \right)^T\left(\begin{array}{cccc} 
Z_2^{-1} & 0 \\
0 & Z_2^{-1} 
\end{array} \right)^{-1}\left(\begin{array}{cccc} 
\frac{1}{2}Z_2^{-1} & 0 \\
0 & -\frac{1}{2}Z_2^{-1}\end{array} \right) 
\end{eqnarray}
are positive definite for $t\in[t_0,t_0+T]$. 

The phase condition ${\rm Re}\,M\succ 0$ guarantees rapid decay of the exponential, while the amplitude $F_{\alpha\beta}(u,v,t;\hbar)F_{\alpha'\beta'}(u',v',t;\hbar)$ exhibits algebraic growth in $(u,v,u',v')$ for $t\in[t_0,t_0+T]$, by linearity of $\kappa$ and by estimate (\ref{eq:estimate}). Thus, the oscillatory integral in (\ref{eq:norm2}) is well defined. Finally, as $F_{\alpha\beta}(t;\hbar)$ is regular in $\hbar$, we conclude that 
\begin{equation}
\Big\|\Big(i\hbar\,\frac{\partial}{\partial t}-\widehat H\Big)G^Z_{(q,p)}(\cdot ,t;\hbar)\Big\|_{L^2}\leq C(q,p,t_0,T)\,\hbar^{3/2}
\end{equation}
where, for some $\hbar_0>0$ 
\begin{eqnarray}
C(q,p,t_0,T)=\frac{\pi^{-d}}{{\rm det}\,Z_2(t)}\,{\rm sup}_{\hbar\in\,]0,\hbar_0[\,}\,\Bigg(\sum_{|\alpha|+|\beta|=3,\,|\alpha'|+|\beta'|=3}\frac{1}{\alpha!\beta!\alpha'!\beta'!}\\ \nonumber
\times \int F_{\alpha\beta}(u,v,t;\hbar)F_{\alpha'\beta'}(u',v',t;\hbar)\,\exp\Big(-(u,v,u',v')\cdot M(t)(u,v,u',v')\Big)\,dudvdu'dv'\Bigg)^{1/2} \ .
\end{eqnarray}

Thus, we have proven that given that the parameters $(q_t,p_t)$, $S(t)$, $Z(t)$ and $a(t)$ satisfy the characteristic system, the propagated Gaussian wave packet $G^Z_{q,p}(t)$ is an asymptotic solution of (\ref{eq:cauchyproblem}) in the sense of (\ref{eq:asol}).

As far as the existence of the solutions of the characteristic system is concerned, their boundedness and uniqueness, these follow from the fact that the corresponding initial value problems are satisfied and by the existence and uniqueness theorem for differential equations, given the smoothness of $H$. Their particular form, as provided, follows from definition and by substitution. However, a comment is required on the well definedness of the dynamical representations (\ref{eq:Z}) and (\ref{eq:a}) of the anisotropy matrix $Z(t)$ and the amplitude $a(t)$. 

We show that the anisotropy matrix, which is well defined, is uniquely expressed in terms of the variational matrices by the relation
\begin{equation}
Z=BA^{-1} \ .
\end{equation}
As ${\rm det}\,A(t_0)$ is non-zero and smooth, there exists some $\varepsilon>0$, such that for $t\in[t_0,t_0+\varepsilon]$ the matrix $BA^{-1}$ is well defined. By the variational system, for $t\in [t_0,t_0+\varepsilon]$, $BA^{-1}$ satisfies the matrix Riccati equation
\begin{eqnarray}
\frac{d}{dt}(BA^{-1})=\frac{dB}{dt}A^{-1}-BA^{-1}\frac{dA}{dt}A^{-1}=\Big(-H_{qq}\,A-H_{qp}\,B\Big)\,A^{-1}-BA^{-1}\Big(H_{pq}\,A+H_{pp}\,B\Big)A^{-1}\\ \nonumber
=-BA^{-1}H_{pp}\,BA^{-1}-BA^{-1}H_{pq}-H_{qp}\,BA^{-1}-H_{qq}
\end{eqnarray}
and also the initial condition, $B(t_0)A(t_0)^{-1}=iI$. By virtue of uniqueness of the solutions of the Cauchy problem of the matrix Riccati equation, $BA^{-1}$ is well defined for $t\in[t_0,t_0+T]$ and $BA^{-1}=Z$ in that interval. Thus, we incur the dynamical representation 
\begin{equation}
Z(t)=\Big(\frac{\partial p_t}{\partial q}+i\frac{\partial p_t}{\partial p}\Big)\Big(\frac{\partial q_t}{\partial q}+i\frac{\partial q_t}{\partial p}\Big)^{-1} \ .
\end{equation}

In turn, by the transport equation, as $A$ is invertible and $a$ is non-zero, we have, by the variational system, 
\begin{eqnarray}
a(t)^2=\exp\Bigg(-\int\displaylimits_{t_0}^t{\rm tr}\Big(H_{pp}Z+H_{pq}\Big)\,d\tau\Bigg)=\exp\Bigg(-\int\displaylimits_{t_0}^t{\rm tr}\Big(H_{pp}B+H_{pq}A\Big)A^{-1}\,d\tau\Bigg)\\ \nonumber 
=\exp\Bigg(-\int\displaylimits_{t_0}^t{\rm tr}\,\frac{dA}{dt}A^{-1}\,d\tau\Bigg)=\frac{1}{{\rm det}\,A(t)}
\end{eqnarray}
and so 
\begin{equation}
a(t)=\Big({\rm det}\,\Big(\frac{\partial q_t}{\partial q}+i\frac{\partial q_t}{\partial p}\Big)\Big)^{-1/2}
\end{equation}
where the branch of the square root is that for which the initial value $a(t_0)=1$ is reached continuously.
\end{proof}

\bigskip

\begin{prop}
\textit{The anisotropic semi-classical Gaussian wave packet $G^Z_{(q,p)}(t)$ acquires the content of the semi-classical correspondent of the pure classical state $(q_t,p_t)$ \cite{NSS}, in the sense that, for any $t\geq t_0$ and $(q,p)\in\mathbb{R}^{2d}$ it satisfies:}

\textit{1. the expectation formulas for the position and momentum operators, for $j=1,\ldots,d$,}
\begin{equation}
\langle \widehat q_j\rangle_{G^Z}=\langle G^Z,\widehat q_j\,G^Z\rangle=q_{t\,j} \ , \ \ \langle \widehat p_j\rangle_{G^Z}=\langle G^Z,\widehat p_j\,G^Z\rangle=p_{t\,j}
\end{equation}

\textit{2. the minimal variance conditions, for $j=1,\ldots,d$,} 
\begin{equation}
(\Delta_{G^Z} \widehat q_j)^2=\langle \widehat q_j^{\,2}\rangle_{G^Z}-\langle \widehat q_j\rangle_{G^Z}^2=\frac{\hbar}{2} \ , \ \ (\Delta_{G^Z} \widehat p_j)^2=\langle \widehat p_j^{\,2}\rangle_{G^Z}-\langle \widehat p_j\rangle_{G^Z}^2=\frac{\hbar}{2}
\end{equation}
\textit{saturating the Heisenberg inequality, $\Delta_{G^Z} \widehat q_j \ \Delta_{G^Z} \widehat p_j=\frac{\hbar}{2}$.}
\end{prop}
 
\begin{proof}
By direct integration, both points follow directly the fact that, for $j=1,\ldots,d$ we have 
\begin{equation}
\widehat q_j \,G^Z_{(q,p)}(x,t;\hbar) = x_j\, G^Z_{(q,p)}(x,t;\hbar) \ , \ \ \widehat p_j \,G^Z_{(q,p)}(x,t;\hbar) =\Big(p_{t\,j}+\sum_{k=1}^dZ(t)_{jk}(x_k-q_{t\,k})\Big)G^Z_{(q,p)}(x,t;\hbar) \ .
\end{equation}

\end{proof}

\section{Variational Matrix Representation of Canonical Relations}

We develop a representation of the underlying invariance relations pertaining to the characteristic system of the Cauchy problem in terms of algebraic relations of the \textit{variational matrices.} The basic aspects of this approach were introduced by Hagedorn \cite{Hag}.

\bigskip

\begin{thm} \label{VMI}\textit{For $t\geq t_0$ and $(q,p)\in\mathbb{R}^{2d}$, the following relations on the position and momentum variational matrices, $A=A(q,p,t_0,t)$ and $B=B(q,p,t_0,t)$, hold}
\begin{eqnarray}
A^TB-B^TA=0\\ \nonumber
\bar AB^T-A\bar B^T=2iI \ , \ \ A\bar A^T-\bar AA^T=0 \ , \ \ B\bar B^T-\bar BB^T=0 \ , \ \ A^*B-B^*A=2iI \\ \nonumber
{\rm Im}\,Z=(A A^*)^{-1} \ , \ \ {\rm Im}\,Z^{-1}=(B B^*)^{-1} \ .
\label{eq:relations}
\end{eqnarray}
\end{thm}

\begin{proof}
The relations between the position and momentum variational matrices represent the symplectic invariance of the Hamiltonian flow and the the symmetry of the anisotropy matrix. 

The anisotropy matrix is symmetric, $Z^T=Z$, as the transpose matrix $Z^T$ satisfies the initial value problem for the matrix Riccati equation for $Z$, as $H_{qq}^T=H_{qq}$, $H_{pp}^T=H_{pp}$, $H_{qp}^T=H_{pq}$ and $(iI)^T=iI$. We proceed to express the above invariances in terms of the variational matrices. 

Symmetry of the anisotropy matrix is expressed as $(BA^{-1})^T=BA^{-1}$, leading to 
\begin{equation}
A^TB-B^TA=0 \ .
\end{equation}

Symplectic invariance is expressed as 
\begin{equation}
\Big(\frac{\partial(q_t,p_t)}{\partial (q,p)}\Big)^TJ\,\frac{\partial(q_t,p_t)}{\partial (q,p)}=J
\end{equation} 
which leads to the Poisson and Lagrange canonical relations \cite{Arn}; in particular, for the canonical co-ordinates $(q_t,p_t)$ and $(q,p)$, respectively, for $j,k=1,\ldots,d$
\begin{eqnarray}
\{q_{t\,j},p_{t\,k}\}=\delta_{jk} \ , \ \ \{q_{t \, j},q_{t \, k}\}=0 \ , \ \ \{p_{t \, j},p_{t \, k}\}=0 \\ \nonumber
[\![q_j,p_k]\!]=\delta_{jk} \ , \ \ [\![q_j,q_k]\!]=0 \ , \ \ [\![p_j,p_k]\!]=0
\end{eqnarray}
where the Poisson brackets of smooth functions $f,g:\mathbb{R}^{2d}\rightarrow \mathbb{R}$, in canonical co-ordinates $(q,p)$, are \cite{Arn}
\begin{equation}
\{f,g\}:=\sum_{l=1}^d\Big(\frac{\partial f}{\partial q_l}\frac{\partial g}{\partial p_l}-\frac{\partial f}{\partial p_l}\frac{\partial g}{\partial q_l}\Big)
\end{equation}
and the Lagrange brackets of $( q_j, p_k)$, for some $j,k=1,\ldots,d$, with respect to the canonical co-ordinates $(q_t,p_t)$ are \cite{Arn} 
\begin{equation}
[\![ q_j, p_k]\!]:=\sum_{l=1}^d\Big(\frac{\partial q_{t\,l}}{\partial q_j}\frac{\partial p_{t\,l}}{\partial p_k}-\frac{\partial q_{t\,l}}{\partial p_k}\frac{\partial p_{t\,l}}{\partial q_j}\Big) \ .
\end{equation}

By inverting relations (\ref{eq:vmatrices}) we obtain a representation of the variations with respect to the initial data of the dynamics in terms of the variational matrices
\begin{equation}
\frac{\partial q_t}{\partial q}=\frac{A+\bar A}{2} \ , \ \ \frac{\partial q_t}{\partial p}=\frac{A-\bar A}{2i} \ , \ \ \frac{\partial p_t}{\partial q}=\frac{B+\bar B}{2} \ , \ \ \frac{\partial p_t}{\partial p}=\frac{B-\bar B}{2i}
\end{equation}
we express the Poisson and Lagrange canonical relations, in terms of the variational matrices; 

For the canonical Poisson relations, $\{q_t,p_t\}=I$, we deduce the relations
\begin{equation}
\bar AB^T-A\bar B^T=2iI
\end{equation}
while for $\{q_t,q_t\}=0$ and $\{p_t,p_t\}=0$ we deduce the relations
\begin{equation}
\bar AA^T-A\bar A^T=0 \ , \ \ \bar BB^T-B\bar B^T=0 \ .
\end{equation} 

For the canonical Lagrange relations, $[\![q,p]\!]=I$ along with $[\![q,q]\!]=0$ and $[\![p,p]\!]=0$, we deduce the relations 
\begin{equation}
A^*B-B^*A=2iI \ .
\end{equation}

Finally, from $A^*B-B^*A=2iI$ and $Z=BA^{-1}$, we deduce the relations 
\begin{eqnarray}
{\rm Im}\,Z=(A A^*)^{-1} \ , \ \ {\rm Im}\,Z^{-1}=(BB^*)^{-1} \ . 
\label{eq:sqrts}
\end{eqnarray}
\end{proof}

\bigskip

\begin{prop}
\textit{The initial value problem for the matrix Riccati equation in (\ref{eq:charsyst}) defines a smooth automorphism on the Siegel half-space \cite{Zel,Fre}}
\begin{equation}
\Sigma_d=\{Z\in\mathbb{C}^{d\times d}\,|\,Z^T=Z,\,{\rm Im}\,Z\succ 0\} \ .
\end{equation}
\end{prop}

\begin{proof}
The fact that the anisotropy matrix is symmetric, $Z^T=Z$, is established in the proof of the theorem above. From relation (\ref{eq:sqrts}) follows that $({\rm Im}\,Z)^{-1}$ is positive definite, and thus its inverse exists and is positive definite. Smoothness follows from the smoothness of the co-efficients of the matrix Riccati equation. Finally, ${\rm Im}\,Z(t)$ remains non-singular for finite time, i.e., $Z(t)$ is bounded away from $\partial\Sigma_d$.
\end{proof}

\bigskip

\begin{prop}
\textit{The relation between the anisotropy matrix $Z$ and the variational matrices $A$, $B$ is unique and established as follows: the anisotropy matrix is uniquely expressed in terms of the variational matrices by}
\begin{equation}
Z=BA^{-1}
\end{equation}
\textit{while the variational matrices, respectively, are uniquely expressed in terms of the anisotropy matrix by}
\begin{equation}
A=({\rm Im}\,Z)^{-1/2} \ , \ \ B=({\rm Im}\,Z^{-1})^{1/2} \ .
\end{equation}
\end{prop}

\begin{proof} The proof of the first relation is given in the proof of the previous theorem. For the second pair of relations, as $Z\in\Sigma_d$ we have that ${\rm Im}\,Z\succ 0$ and subsequently $({\rm Im}\,Z)^{-1}\succ 0$; by the uniqueness of positive definite square roots of positive definite matrices, there exists a unique positive definite matrix $\tilde A\in\mathbb{C}^{d\times d}$ such that $({\rm Im}\,Z)^{-1}=\tilde A\tilde A^*$. However, given relation (\ref{eq:sqrts}) $AA^*=({\rm Im}\,Z)^{-1}$, the unique positive square root of $({\rm Im}\,Z)^{-1}$ is $\tilde A=A$. The fact that $B=({\rm Im}\,Z^{-1})^{-1/2}$ follows from $B=ZA$.
\end{proof}

\bigskip

\begin{rem}
The anisotropic Gaussian wave packet (\ref{eq:agwp}) is invariant under the right action of the gauge group ${\rm SU}(d)$ on the variational matrices, in the sense that for $U\in{\rm SU}(d)$ the action $(A,B)\mapsto (AU,BU)$ preserves $Z$, as $(BU)(AU)^{-1}=BUU^{-1}A^{-1}=Z$ as well as $a(t)=({\rm det}\,A)^{-1/2}\mapsto ({\rm det}\,AU)^{-1/2}=({\rm det}\,A)^{-1/2}$ \cite{Hag}. An immediate consequence is the gauge invariance of the relations (\ref{eq:relations}). This gauge invariance is a manifestation of the symplectic invariance of the variational system under; the action of the element $U=U_1+iU_2\in{\rm SU}(d)$ on the variational matrices, for real $U_1,U_2$, is equivalent to the action of ${\rm Sp}(2d)$ on $\mathbb{R}^{2d}$
\begin{equation}
\left(\begin{array}{ccc} q \\ p \end{array} \right)\mapsto \left(\begin{array}{ccc} U_1 & U_2 \\ -U_2 & U_1 \end{array} \right)\left(\begin{array}{ccc} q \\ p \end{array} \right)
\end{equation}
since $U_1^TU_1+U_2^TU_2$ and $U_1^TU_2-U_2^TU_1=0$.
\end{rem}

\section{Inference of the Van Vleck Formula}

The \textit{Van Vleck formula} \cite{VVl} is a weak semi-classical expansion of the Schr\"{o}dinger propagator kernel, which reads
\begin{equation}\label{vanvleck}
K(x,y,t_0,t;\hbar)\sim\Big(\frac{1}{2\pi i\hbar}\Big)^{d/2}\sum_{r=1}^{N}\sqrt{\Big|\det\,\frac{\partial p}{\partial q_t}(t;y,p_r(y,x,t_0,t))\Big|}\,e^{\frac{i}{\hbar} S(y,p_r(y,x,t_0,t),t_0,t)-\frac{ i\pi }{2}m_r} \ , \ \ \hbar\rightarrow 0^+ \ .
\end{equation}
Here, $N=N(x,y,t_0,t)$ is the number of orbits of the Hamiltonian flow \cite{Arn} emanating from point $(q,p)$ at time $t_0$, terminating at point $(q_t,p_t)$ at time $t\geq t_0$, such that $q=y$ and $q_t=x$; for $r=1,\ldots,N$, $p_r=p_r(y,x,t_0,t)$ are the admissible initial momenta of the orbit and $m_r$ is the Maslov index of the orbit; $S=S(q,p,t_0,t)$ is the action of the orbit emanating from point $(q,p)$ at time $t_0$ \cite{Arn}. We assume that $N$ is finite, but, in general, an upper bound for $N$ cannot be found \cite{SMM}.The above formula is typically derived by means of a formal stationary phase asymptotic expansion of the Feynmann path integral representation of the kernel of the Schr\"{o}dinger propagator \cite{VVl,FeHi,BeSh}.

In this section we infer the Van Vleck formula on the basis of the Anisotropic Gaussian Approximation. A derivation of the Van Vleck formula in this frame-work, though in a different way, is given by Billy and Robert \cite{BiRo} and Blair \cite{Bla}. 

\bigskip

\begin{defn} 
\textit{We define the} Anisotropic Gaussian semi-classical Schr\"{o}dinger propagator \textit{as the operator on $L^2(\mathbb{R}^d)$}
\begin{equation}
U^Z(t_0,t):=\Big(\frac{1}{2\pi\hbar}\Big)^{d}\int G^Z_{(q,p)}(t)\langle G_{(q,p)},\cdot\rangle \,dqdp 
\end{equation}
\textit{where $G^Z_{(q,p)}(t)$ is given in (\ref{eq:agwp}). In particular, it is characterized by the action $U^Z(t_0,t)G_{(q,p)}=G^Z_{(q,p)}(t)$. It is a semi-classical Fourier Integral Operator, acting as \cite{Rob1,PaUr}}
\begin{equation}
U^Z(t_0,t)\psi(x)=\int K^Z(x,y,t_0,t;\hbar)\psi(y)\,dy
\end{equation}
\textit{its kernel given by the oscillatory integral distribution \cite{KaMa}}
\begin{equation}
K^Z(x,y,t_0,t;\hbar)=\Big(\frac{1}{2\pi\hbar}\Big)^{d}\int\bar G_{(q,p)}(y;\hbar) G^Z_{(q,p)}(x,t;\hbar)\,dqdp \ .
\end{equation}
\end{defn}

\bigskip

The operator $U^Z(t_0,t)$ has been studied by Paul and Uribe \cite{PaUr}, Rousse and Swart \cite{RoSw}, Robert \cite{Rob2} and Blair \cite{Bla}. We characterize it as a semi-classical Fourier Integral Operator, in the sense that it semi-classically approximates the Schr\"{o}dinger propagator for short times, yet, with an oscillatory distribution operator kernel whose phase function is complex valued and quadratic in $x$. Such generalizations of conventional Fourier Integral Operators for the solution of the Schr\"{o}dinger equation have been studied by Laptev and Sigal \cite{LaSi}.\

We proceed with a result deriving from an important theorem by Nazaikinskii, Oshmyan, Sternin and Shatalov in \cite{NOSS}, similar to the one of Melin and Sj\"{o}strand \cite{MeSj}.

\bigskip

\begin{thm} \label{SPT} \textit{Consider the oscillatory integral}
\begin{equation}
I_{\varphi,\Phi}(w;\hbar)=\Big(\frac{1}{2\pi\hbar}\Big)^{d}\int \varphi(X)\,e^{\frac{i}{\hbar}\Phi(w,X)}\,dX
\end{equation}
\textit{where $\varphi\in C_0^\infty(\mathbb{R}^{2d},\mathbb{C})$ and $\Phi\in C^{\infty}(\mathbb{R}^m\times \mathbb{R}^{2d},\mathbb{C})$, where $\Phi$ possesses an everywhere non-negative imaginary part on ${\rm supp}\,\varphi$, ${\rm Im}\,\Phi(w,X)\geq 0$, so that the equations}
\begin{equation}
{\rm Im}\,\Phi(w,X)=0 \ , \ \ \frac{\partial \Phi}{\partial X}(w,X)=0
\end{equation}
\textit{have $N$ discrete solutions on ${\rm supp}\,\varphi$ for given $w\in\mathbb{R}^m$, denoted $X=X_r(w)$, for $r=1,\ldots,N$, while ${\rm det}\,\frac{\partial^2\Phi}{\partial X^2}(w,X_r(w))\neq 0$. Then, these solutions are real and the following estimate holds}
\begin{equation}
I_{\varphi,\Phi}(w;\hbar)=\sum_{r=1}^N\frac{\varphi(X_r(w))}{\sqrt{{\rm det}\,-\frac{\partial^2\Phi}{\partial X^2}(w,X_r(w))}}\,e^{\frac{i}{\hbar}\Phi(w,X_r(w))}\Big(1+o(\hbar)\Big) \ , \ \ \hbar\rightarrow 0^+
\end{equation}
\textit{where $\sqrt{\cdot}$ is the principal branch of the square root function.}
\end{thm}

\begin{proof}
The proof follows trivially as an immediate extension of that of Nazaikinskii, Oshmyan, Sternin and Shatalov in \cite{NOSS}. The supplemented consists of taking, for each $w\in\mathbb{R}^m$, appropriate neighborhood of $X_r(w)$ contained in ${\rm supp}\,\varphi$, for $r=1,\ldots,N,$ each neighborhood unique to each solution, and repeating the proof for each neighborhood.
\end{proof}

\bigskip

We give the main result of this section, noting that it has been produced, on different premisses and methodology, by Billy and Robert \cite{BiRo} and Blair \cite{Bla}.

\bigskip

\begin{thm} \textit{Let $H$ be as in (\ref{eq:estimate}), with the additional assumption that the generated flow develops no caustics in the given time interval, in the sense that ${\rm det}\,\frac{\partial q_t}{\partial p}(q,p)\neq 0$ for $t_0\leq t\leq t_0+T$. Also, for $n\in\mathbb{N}$, let the sequence of cut-off functions $\chi_n\in C^\infty_0(\mathbb{R},\mathbb{C})$ have connected support, containing the ball $B_{\rho_n}(0)$, for some increasing diverging real sequence $\{\rho_n\}$, taking the value 1 away from its boundary, and satisfying the point-wise condition $\lim_{n\rightarrow+\infty}\chi_n(q,p)=1$ for all $(q,p)\in\mathbb{R}^{2d}$. We define the approximate propagator $U^Z_{\chi_n}(t_0,t)$, with kernel}
\begin{equation}
K^Z_{\chi_n}(x,y,t_0,t;\hbar)=\Big(\frac{1}{2\pi\hbar}\Big)^d\int \chi_n(q,p)\bar G_{(q,p)}(y;\hbar) G^Z_{(q,p)}(x,t;\hbar)\,dqdp \ .
\end{equation}
\textit{Then, the kernel $K^Z_{\chi_n}$ satisfies the Van Vleck formula, in the sense that, for given $x,y,t_0,t$ and for large enough $n$, we have}
\begin{eqnarray}
K^Z_{\chi_n}(x,y,t_0,t;\hbar)=\Big(\frac{1}{2\pi i\hbar}\Big)^{d/2}\sum_{r=1}^{N}\chi_n\Big(y,p_r(y,x,t_0,t)\Big)\\ \nonumber 
\times \sqrt{\Big|\det\,\frac{\partial p}{\partial q_t}(t;y,p_r(y,x,t_0,t))\Big|}\,e^{\frac{i}{\hbar} S(y,p_r(y,x,t_0,t),t_0,t)-\frac{ i\pi }{2}m_r}\Big(1+o_n(\hbar)\Big) \ , \ \ \hbar\rightarrow 0^+
\end{eqnarray}
where the $N$ momenta $p_r(y,x,t_0,t)$ are the admissible initial momenta of the orbit emanating from point $(q,p)$ at time $t_0$, terminating at point $(q_t,p_t)$ at time $t\geq t_0$, such that $q=y$ and $q_t=x$.
\end{thm}

\begin{proof}
Expressing the kernel $K^Z_{\chi_n}$ in the form of theorem (\ref{SPT}), we have 
\begin{equation}
K^Z_{\chi_n}(x,y,t_0,t;\hbar)=\Big(\frac{1}{2\pi\hbar}\Big)^{3d/2}\int \varphi_n(q,p,t_0,t)\,e^{\frac{i}{\hbar}\Phi(x,y,q,p,t_0,t)}\,dqdp
\end{equation}
where the phase is 
\begin{eqnarray}
\Phi(x,y,q,p,t_0,t)= S(q,p,t_0,t)+p_t\cdot (x-q_t)+\frac{1}{2}(x-q_t)\cdot Z(q,p,t_0,t)(x-q_t)\\ \nonumber
-p\cdot (y-q)+\frac{i}{2}|y-q|^2
\end{eqnarray}
and the amplitude is
\begin{equation}
\varphi_n(q,p,t_0,t)=2^{d/2}\chi_n(q,p)\Big({\rm det}\,{\rm Im}\,Z(q,p,t_0,t)\Big)^{1/4}
\end{equation}
the later by virtue of the identity 
\begin{equation}
({\rm det}\,A)^{-1/2}=({\rm det\,Im}\,Z)^{1/4}
\label{eq:dets}
\end{equation}
 which derives from (\ref{eq:sqrts}). 

Both phase and amplitude satisfy the smoothness conditions of theorem (\ref{SPT}). In addition, we have that 
\begin{equation}
{\rm Im}\,\Phi(t)= \frac{1}{2}(x-q_t)\cdot {\rm Im}\,Z(t)(x-q_t)+\frac{1}{2}|y-q|^2\geq 0
\end{equation}
everywhere, as ${\rm Im}\,Z\succ 0$, so that the zero-level set of the imaginary part of the phase $\Phi$, for fixed $x$, $y$ and $t>t_0$, is 
\begin{equation}
\{(q,p)\in\mathbb{R}^{2d}\,|\,q=y,\,\pi g^t(q,p)=x\}
\end{equation}
where $\pi:\mathbb{R}^{2d}\rightarrow \mathbb{R}^{d}$ is the canonical projection from phase space onto configuration space. Given the smoothness of $H$, the zero-level set is either finite or empty. For $(q,p)$ belonging to the zero-level set, we trivially have 
\begin{equation}
\frac{\partial\Phi}{\partial q}=0 \ , \ \ \frac{\partial\Phi}{\partial p}=0
\end{equation} 
by virtue of the property of the action
\begin{equation}
\frac{\partial S}{\partial q}=-p+p_t\cdot\frac{\partial q_t}{\partial q} \ , \ \ \frac{\partial S}{\partial p}=p_t\cdot\frac{\partial q_t}{\partial p} \ . 
\end{equation}

The zero-level set consists of all initial phase space points whose orbit project to trajectories on configuration space joining points $y$ and $x$ between time instances $t_0$ and $t$, respectively. The equation $x=q_t(q,p)$ for $q=y$, becomes the equation for the possible initial momenta $p$ for which a trajectory satisfies the boundary conditions $q=y$ and $q_t=x$. 

As the Hamiltonian is smooth in $(q,p)$ and continuous in $t$, the final position is smooth in its dependence on the initial position $q$. Thus, given that for $t_0\leq t\leq t_0+T$ we have ${\rm det}\,\frac{\partial q_t}{\partial p}\neq 0$, the equation $q=q_t(y,p)$ possesses countably many solutions, $p_r=p_r(y,x,t_0,t)$ for $r=1,\ldots,N$, for some $N=N(x,y,t_0,t)$. 

By direct computation, for $(q,p)=(y,p_r(y,x,t_0,t))$, for any $r=1,\ldots,N$, the Hessian blocks of the phase are given by 
\begin{eqnarray}
\Phi_{qq}=iI-\Big(\frac{\partial p_t}{\partial q}\Big)^T \frac{\partial q_t}{\partial q}+\Big(\frac{\partial q_t}{\partial q}\Big)^T Z\,\frac{\partial q_t}{\partial q} \\ \nonumber
\Phi_{pp}=-\Big(\frac{\partial p_t}{\partial p}\Big)^T \frac{\partial q_t}{\partial p}+\Big(\frac{\partial q_t}{\partial p}\Big)^T Z\,\frac{\partial q_t}{\partial p}\\ \nonumber
\Phi_{qp}=-\Big(\frac{\partial p_t}{\partial q}\Big)^T \frac{\partial q_t}{\partial p}+\frac{1}{2}\Big(\frac{\partial q_t}{\partial q}\Big)^T Z\,\frac{\partial q_t}{\partial p}+\frac{1}{2}\Big(\frac{\partial q_t}{\partial p}\Big)^T Z\,\frac{\partial q_t}{\partial q}\\ \nonumber
\Phi_{pq}=-\Big(\frac{\partial q_t}{\partial p}\Big)^T \frac{\partial p_t}{\partial q}+\frac{1}{2}\Big(\frac{\partial q_t}{\partial p}\Big)^T Z\,\frac{\partial q_t}{\partial q}+\frac{1}{2}\Big(\frac{\partial q_t}{\partial q}\Big)^T Z\,\frac{\partial q_t}{\partial p}
\end{eqnarray}
while the determinant of the Hessian matrix on the zero-level set is 
\begin{equation}
{\rm det\,Hess}_{(q,p)}\Phi={\rm det}\,\Bigg(\frac{2}{i}\Big(\frac{\partial q_t}{\partial q}+i\frac{\partial q_t}{\partial p}\Big)^{-1}\frac{\partial q_t}{\partial p}\Bigg)
\end{equation}
which is non-singular, as the matrices $\frac{\partial q_t}{\partial q}+i\frac{\partial q_t}{\partial p}=A(t)$ and $\frac{\partial q_t}{\partial p}$ are non-singular within the given time interval.

Thus, we have
\begin{eqnarray}
\frac{\varphi_n(y,p_r(y,x,t_0,t),t_0,t)}{\sqrt{{\rm det}-{\rm Hess}_{(q,p)}\Phi(x,y,y,p_r(y,x,t_0,t),t_0,t)}}=2^{d/2}\chi_n(y,p_r(y,x,t_0,t))\\ \nonumber
\Big({\rm det}\,{\rm Im}\,Z(y,p_r(y,x,t_0,t),t_0,t)\Big)^{1/4}\Bigg\{{\rm det}\,\Bigg(-\frac{2}{i}\Big(\frac{\partial q_t}{\partial q}+i\frac{\partial q_t}{\partial p}\Big)^{-1}\frac{\partial q_t}{\partial p}\Bigg)\Bigg\}^{-1/2}\\ \nonumber
=i^{-d/2}\,\chi_n(y,p_r(y,x,t_0,t)){\sqrt{{\rm det}\,\frac{\partial p}{\partial q_t}(y,p_r(y,x,t_0,t))}}
\end{eqnarray}
by virtue of relation (\ref{eq:dets}).

Assuming $n$ is large enough so that ${\rm supp}\,\varphi_n(\cdot ,t_0,t)$ contains the zero-level set and that $T=O(1)$, conditions of theorem (\ref{SPT}) are met for each $t_0$ and $t\in[t_0,t_0+T]$, so that we obtain
\begin{eqnarray}
K_{\chi_n}^Z(x,y,t_0,t;\hbar)=\Big(\frac{1}{2\pi i\hbar}\Big)^{d/2}\sum_{r=1}^{N}\chi_n(y,p_r(y,x,t_0,t))\sqrt{\Big|\det\,\frac{\partial p}{\partial q_t}(y,p_r(y,x,t_0,t))\Big|}\\ \nonumber 
\times\, e^{\frac{i}{\hbar}S(y,p_r(y,x,t_0,t),t_0,t)-\frac{i\pi}{2}m_r}\Big(1+o_n(\hbar)\Big) \ , \ \ \hbar\rightarrow 0^+ \ .
\end{eqnarray}
In the above, the index $r=1,\ldots,N$ enumerates the orbits emanating from point $(q,p)$ at time $t_0$ with momentum $p_r(y,x,t_0,t)$, terminating at point $(q_t,p_t)$ at time $t\geq t_0$, such that $q=y$ and $q_t=x$, while 
\begin{equation}
m_r={\rm ind}\Big(\Big(\frac{\partial q_t}{\partial q}+i\frac{\partial q_t}{\partial p}\Big)^{-1}\frac{\partial p}{\partial q_t}\Big):=\sum_{\lambda}{\rm Arg}(\lambda)
\end{equation}
for $\lambda\in{\rm Spec}\Big(\Big(\frac{\partial q_t}{\partial q}+i\frac{\partial q_t}{\partial p}\Big)^{-1}\frac{\partial p}{\partial q_t}\Big)$, i.e., the index of the monodromy matrix, the excess of its positive over negative eigenvalues.
\end{proof}

\begin{rmk}
As $n \to +\infty$, $K_{\chi_n}^Z$ clearly converges to $K^Z$, and we formally expect that the remainder term $o_n(\hbar)$ vanishes, so that $(\ref{vanvleck})$ holds.
\end{rmk}


\begin{thebibliography}{24}

\bibitem{Arn} Arnold V I 1978 \textit{Mathematical Methods of Classical Mechanics} in \textit{Graduate Texts in Mathematics} Vol. \textbf{60} (New York: Springer-Verlag)

\bibitem{BaDa} Babich V M and Danilov Yu P 1969 \textit{Zap. Nauchn. Sem. LOMI} {\bf 15} 47 (in the Russian language)

\bibitem{BBT} Bagrov V G, Belov V V and Trifonov A Yu 1996 \textit{Ann. Phys.} {\bf 246} 231

\bibitem{BeSh} Berezin F A and Shubin M A 1983 \textit{The Schr\"{o}dinger Equation} in \textit{Mathematics and its Applications (Soviet Series)} Vol. \textbf{66} (Dodrecht: Springer Science+Bussiness Media)

\bibitem{BiRo} Bily J M and Robert D 1999 \textit{The Semiclassical Van-Vleck Formula. Application to the Aharonov-Bohm Effect} in \textit{Long Time Behaviour of Classical and Quantum Systems, Series on Concrete and Applicable Mathematics} Vol. {\bf 1} Graffi S and Martinez A (editors) (Singapore: World Scientific)

\bibitem{Bla} Blair M D 2020 \textit{The Van Vleck Formula on Ehrenfest Time Scales and Stationary Phase Asymptotics for Frequency Dependent Phases} arXiv:2012.13034v1

\bibitem{CoFe} Cordoba A and Fefferman C 1978 \textit{Commun. Part. Diff. Eq.} {\bf 3} 979

\bibitem{FeHi} Feynman R P and Hibbs A R 2010 \textit{Quantum Mechanics and Path Integrals} (Mineola: Dover Publications, Incorporated)

\bibitem{Fol} Folland G B 1989 \textit{Harmonic Analysis in Phase Space} (Princeton: Princeton University Press)

\bibitem{Fre} Freitas P J 1999 \textit{On the Action of the Symplectic Group on the Siegel Upper Half Plane} Doctorate Thesis (Chicago: University of Illinois)

\bibitem{Hag} Hagedorn G A 1980 \textit{Commun. Math. Phys.} {\bf 71} 77

\bibitem{Hel} Heller E J 1976 \textit{J. Chem. Phys.} {\bf 64} 63

\bibitem{HeKl} Herman M F and Kluk E 1984 \textit{Chem. Phys.} {\bf 91} 27

\bibitem{HoZh} Horn R A and Zhang F 2005 \textit{The Schur Complement and Its Applications} Zhang F (editor) (New York: Springer)

\bibitem{HHH} Huber D, Heller E J and Harter W G 1987 \textit{J. Chem. Phys.} {\bf 87} 1116

\bibitem{HHL} Huber D, Heller E J and Littlejohn R G 1988 \textit{J. Chem. Phys.} {\bf 89} 2003

\bibitem{KaMa} Karageorge P and Makrakis G 2020 \textit{Asymptotics for the Phase Space Schrödinger Equation} arXiv:2005.08558

\bibitem{KKL} Katchalov A P, Kurylev Y and Lassas M 2001 \textit{Inverse Boundary Spectral Problems} (Boca Raton: Chapman $\&$ Hall/CRC)

\bibitem{LaSi} Laptev A and Sigal I M 2000 \textit{Rev. Math. Phys.} {\bf 12} 749

\bibitem{Lit} Littlejohn R G 1992 \textit{J. Stat. Phys.} {\bf 68} 7

\bibitem{MeSj} Melin A and Sj\"{o}strand J 1975 \textit{Fourier Integral Operators With Complex-Valued Phase Function} in \textit{Lect. Notes Math.} Vol. {\bf 459} (Cham: Springer-Verlag)

\bibitem{NSS} Nazaikinskii V E, Schulze B W and Sternin B Yu 2002 \textit{Quantization Methods in Differential Equations} (New York: Taylor $\&$ Francis)

\bibitem{NOSS} Nazaikinskii V E, Oshmyan V G, Sternin B Yu and Shatalov V E 1981 \textit{Russ. Math. Surv.+} {\bf 36} 93

\bibitem{PaUr} Paul T and Uribe A 1995 \textit{J. Funct. Anal.} {\bf 132} 192

\bibitem{Ral} Ralston J 1983 \textit{MAA Studies in Mathematics} \textbf{23} 206

\bibitem{RDD} Rahimi A, Daraby B and Darvishi Z 2016 \textit{Construction of Continuous Frames in Hilbert Space} arXiv:1606.08981v1

\bibitem{Rob1} Robert D 2007 \textit{Seminaires $\&$ Congr$\grave{e}$s: Partial Differential Equations and Applications} Vol. {\bf 15} 181

\bibitem{Rob2} Robert D 2010 \textit{Rev. Math. Phys.} {\bf 22} 1123

\bibitem{RoSw} Rousse V and Swart T 2009 \textit{Commun. Math. Phys.} {\bf 286} 725

\bibitem{SMM} Sparber C, Markowich P A and Mauser N J 2003 \textit{ Asymptot. Anal.} {\bf 33} 153

\bibitem{TAG} Twareque A S, Antoine J P and Gazeau J P 1993 \textit{Ann. Phys.-New York} {\bf 222} 1

\bibitem{VVl} Van Vleck J H 1928 \textit{Proc. Natl. Acad. Sci. USA.} {\bf 14} 178

\bibitem{Zel} Zelikin M I 1992 \textit{Math. USSR Sb.+} {\bf 73} 341

\end{thebibliography}
\end{document}